\documentclass[11pt]{article}
\usepackage{moriond,epsfig}
\input psfig.tex

\newcommand{\ga}{\raisebox{-0.5ex}{$\,\stackrel{>}{\scriptstyle\sim}\,$}}

\def\be{\begin{equation}}
\def\ee{\end{equation}}
\def\bea{\begin{eqnarray}}
\def\eea{\end{eqnarray}}

\begin{document}
\vspace*{4cm}
\title{Colour--detected clusters in the XMM Large Scale Structure Survey}

\author{S. Andreon$^1$, J. Willis$^2$, H. Quintana$^3$,
I. Valtchanov$^4$, M. Pierre$^5$, F. Pacaud$^5$}

\address{$^1$ INAF--Osservatorio Astronomico di Brera, Milano, Italy \\
$^2$ Department of Physics and Astronomy, University of Victoria, Victoria, 
Canada \\
$^3$ Departamento de Astronom\'\i a y Astrof\'\i sica, Pontificia Universidad 
Cat\'olica de Chile, Santiago, Chile\\
$^4$ Imperial College, London, UK \\
$^5$ CEA/DSM/DAPNIA, Service d'Astrophysique, Gif-sur-Yvette, France\\
}

\maketitle\abstracts{We present first results on the use of a 
colour space filter for detecting galaxy clusters
at cosmological redshifts in the XMM Large Scale Structure Survey.
All clusters studied, but one, are successfully colour--detected in spite
of their large redshift ($0.3<z<1.0$), X--ray selection and intrinsic low
richness ($R=0$ or below). We experimentally show that the cluster redshift can be derived,
with good accuracy, from the colour of the red sequence, at $0.06<z<1$ using
a sample of  about 160 clusters.}

\section{Introduction} 

The XMM--Large Scale Structure (XMM--LSS) is a unique project, currently
gathering a multi-$\lambda$ coverage over a single area of some 6 deg$^2$:
X--ray (XMM), optical/near--infrared (CFHTLS, CTIO, UKIDSS), Mid/Far--infrared
(Spitzer Legacy: SWIRE), Radio continuum (VLA survey), UV (Galex) (Pierre et
al.\cite{Petal04}, Andreon \& Pierre\cite{AP03}). Since the cluster number
density and correlation function depend on key cosmological parameters, such as
the matter density, $\Omega_m$, and, even more sensitively, the present-day
amplitude of density fluctuations on scale of 8 Mpc, $\sigma_8$ (see Refregier
et al.\cite{R02}, Evrard et  al.\cite{E02}), an efficient cluster detection is
very important. Several techniques (X--ray, optical, Sunayev--Zeldovich effect,
lensing) are being used with success to efficiently detect clusters of galaxies out
to $z \approx 1$ and above. The use of many methods should give us an insight
on the selection effects of each technique, and  correspondingly the impact of
the derived cluster characteristics.

\begin{figure}
\begin{minipage}[t]{10truecm}
\psfig{figure=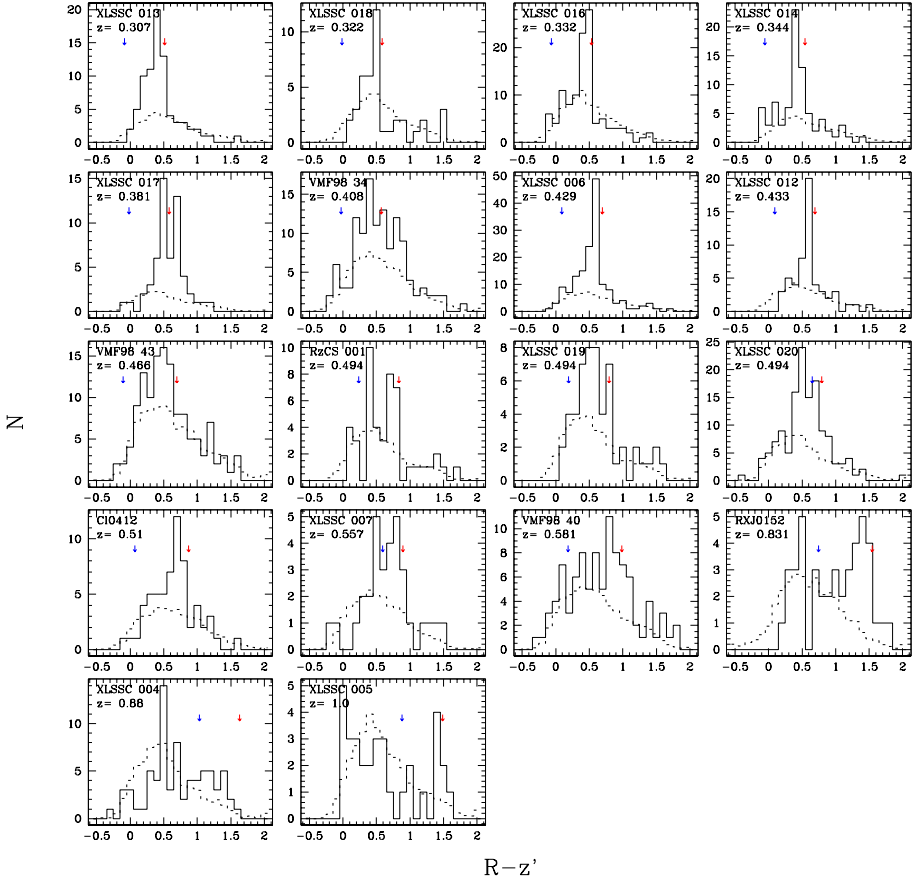,width=12.5truecm}
\end{minipage}
\hskip 2.5 truecm \vbox{\hsize=3.3truecm
\caption{Colour histograms of galaxies in each cluster field. The solid
histogram refers to the cluster field, while the dashed histogram
indicates the colour distribution of the control field normalized to
the cluster field area. The arrows mark the expected
colour range for galaxies at the cluster redshift. Given the
observed overdensity, it is not surprising that 
all clusters, but
XLSSC 007, show a statistically significant overdensity of galaxies
in the cluster line of sight in an optimally chosen colour range. 
Statistically significant peaks outside the colour range are due to another (spectroscopic
confirmed) cluster on the line of sight.
\hfill
\quad
}
}
\end{figure} 

One of the methods used for cluster detection makes resemblance to 
the ''red sequence" method (Gladders \& Yee\cite{GY00}), but 
differs in several
key areas (see Andreon\cite{A03} for details). Both methods 
exploit the observed trend that the majority of galaxies in
clusters display similar colours, while non--cluster
galaxies located along the line--of--sight display considerable
variation of observed colours, both because they are drawn from a
larger interval of redshift and because the field galaxy population at
a given redshift displays a larger variation in colour than a typical
cluster galaxy population (see Fig 1).

Here we focus on 18 clusters at $z\ga0.3$, most of which drawn
from the XMM--LSS survey. Out of the 18 clusters, 16 are X-ray selected.
Observations have been performed in $R$-- and
$z'$--band ($\lambda_c\sim9000${\AA}) at the
Cerro Tololo Inter--American Observatory (CTIO) 4m Blanco telescope
during two observing runs (2000--2001) with the
Mosaic II camera. Mosaic II is a 8k$\times$8k camera with a $36 \times
36$ arcminute field of view. Typical exposure times were 1200 seconds
in $R$ and $2 \times 750$ seconds in $z'$. 
A throughout discussion of most of the results presented here can be
found in Andreon et al.\cite{P4}. When noted, this sample is supplemented
by 140 clusters at $0.06<z<0.30$ colour--detected using SDSS photometry (see Andreon
2003 for details).

\section{Results}

The colour distribution of galaxies brighter than the $z'$--band
completeness limit within each cluster field is displayed in Figure 1 for
the 18 $z \ga 0.3$ clusters.
The colour distribution along the cluster line of sight
is compared to the colour distribution
measured in a 0.3 deg$^2$ control field and
normalized to the cluster area to which it is compared.   
All clusters within the sample display a significant
numerical excess over a limited colour interval (typically $\pm 0.3$
mag), indicating that clusters may be identified effectively 
by methods that employ colour selection to
suppress background galaxy signals.  All clusters, with
the exception of XLSSC 007, were in fact colour--detected at high
significance.

The clusters shown in Fig. 1 display a range of masses (as
determined by either dynamical or X--ray information, or both).  In
particular, XLSSC clusters at $z<0.6$ display X--ray luminosities
comparable to low richness clusters or groups (see Andreon et al.\cite{P4}
and Willis et al.\cite{P3}). Therefore, the
detection of galaxy overdensities in the 3D--space defined by colour
and sky location, at the location of extended X--ray sources, indicates
that such techniques may provide a promising route to confirm the
nature of low mass X--ray selected clusters.

\begin{figure*}
\centerline{%
\psfig{figure=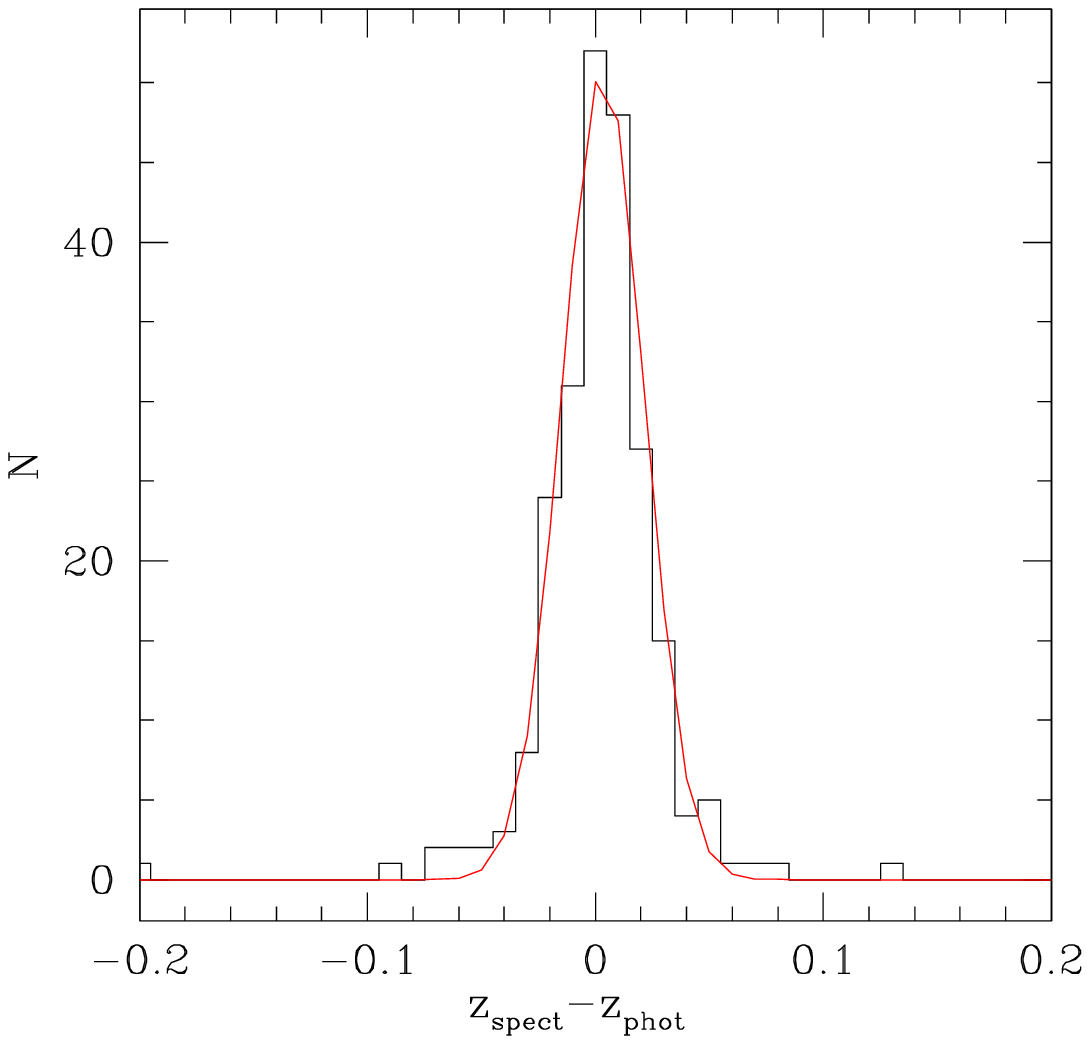,width=7.5truecm}
\hfill
\psfig{figure=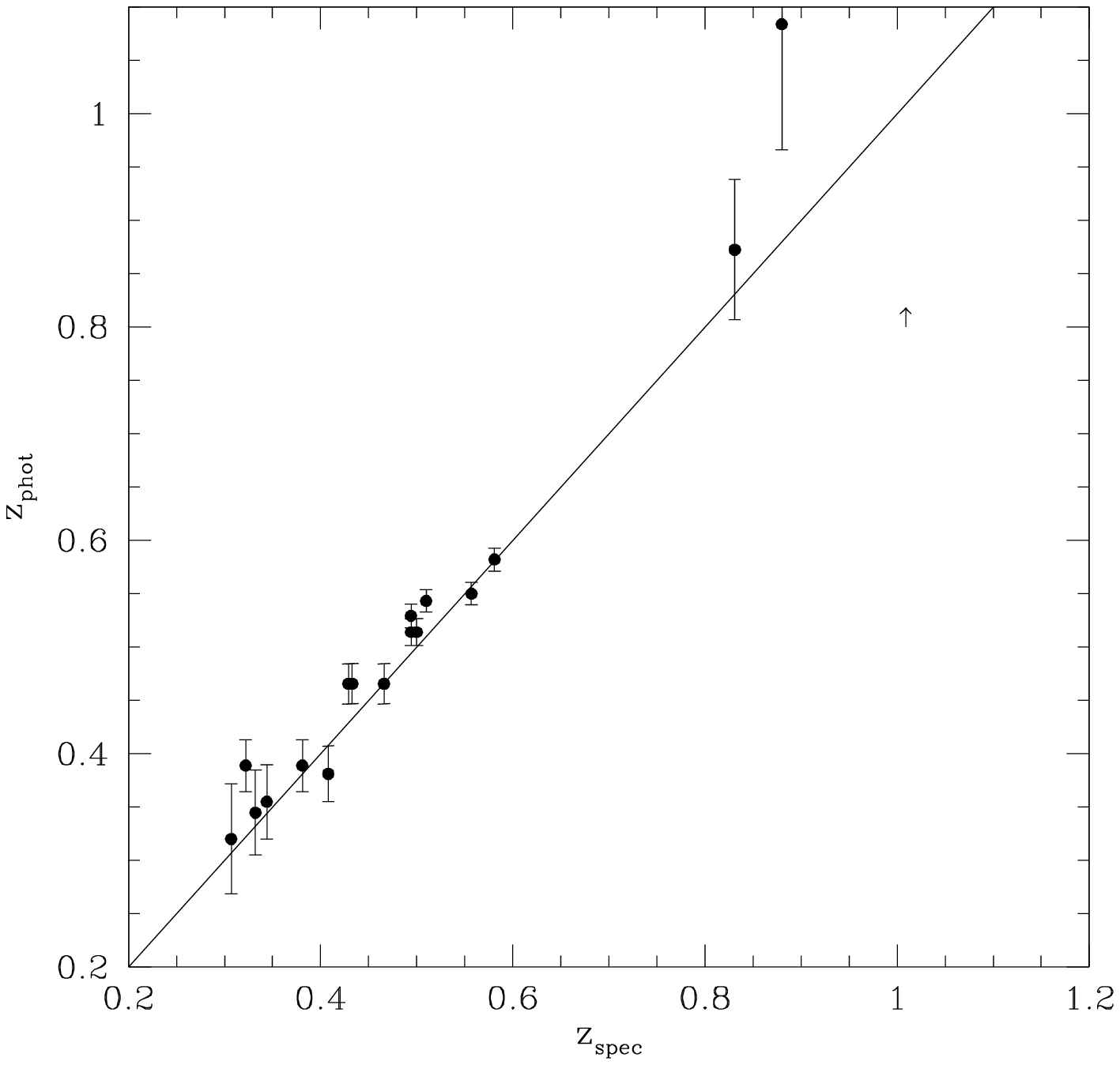,width=7.5truecm}
}
\caption[l]{
{\it Left: } Histogram of the difference between $z_{spectr}$ and
$z_{phot}$ the latter derived from the $g-r$ colour of the red sequence, for
140 clusters in the $0.06<z<0.30$ redshift range (Andreon\cite{A03}). The curve
is a Gaussian of the same scatter (0.018 in z).
{\it Right:} Spectroscopic vs photometric redshift, the
latter derived from the $R-z'$ colour of the red sequence for
18 clusters, mostly in the XMM--LSS survey. \hfill
\quad
}
\end{figure*}

However, the failure to combine X--ray data with colour selected cluster
samples could lead to a number of sources of bias, as colour plus position
selection alone does not constrain the extension in redshift of the identified
structure. A filamentary structure of galaxies seen along the line of sight is,
without spectroscopic data, hard to distinguish from a cluster and both
scenarios can in principle give rise to the ``cluster" detection. Spectroscopic
observations of a sample of colour selected structures are therefore required
to measure the frequency of each type of structure (clusters versus
non--virialised large--scale structure). 

The spectroscopic cluster sample presented in this paper
contains one colour selected cluster undetected in X--ray
(RzCS 011, at $z=0.494$). This system is confirmed spectroscopically
and displays a well--defined mean redshift and distribution of
rest--frame velocities. It is, therefore, a cluster in the sense of
being a gravitationally bound systems of galaxies, although 
undetected in X--rays.  Therefore, colour
selection techniques provide a method to identify clusters displaying
a broad range of X--ray properties, possibly
sampling the cluster mass function deeper than X--ray observations.

Figure 2 shows a measure of the performance of photometric redshifts
derived from the colour of the red sequence.  The right panel of Fig. 2 shows the photometric
vs spectroscopic redshift comparison for 18 clusters shown in Figure 1.
The photometric redshift assumes that the colour of the red sequence
evolves as a passive ``elliptical" galaxy, i.e. following the colour track
of Kodama \& Arimoto\cite{KA97}, as detailed in Andreon et al.\cite{P4}. The agreement is good. Error bars are
larger at the two redshift ends because the photometric redshift accuracy
is proportional to the inverse of the derivative of the colour--redshift relationship, which
is flatter at the two redshift extrema than in between. At these ends,
another choice of filters would be more effective for the photometric
redshift determination, such as those allowed by the upcoming CFHLS and UKIDSS
projects. On the left panel, the $g'-r'$
SDSS colour is used, for 140 colour--detecetd clusters in the $0.06<z<0.30$ redshift
range (Andreon\cite{A03}). The scatter amounts to only 0.018
in redshift, at least three time better than most photometric redshift
estimates. Puddu\cite{Pu01} also show a similar comparison for a small,
but X--ray selected, cluster sample.

The extremely good performance of the red sequence colour as a
redshift indicator is hardly surprising, because of the implicit selection
of one single type of galaxies with a distinctive 4000 \AA \ break
(spectrophotometric bright early--type galaxies) and of the colour homogeneity
of the early--type galaxy class (e.g. Stanford, Eisenhardt,  \& 
Dickinson\cite{SED98}, Kodama et al.\cite{Ketal98},
Andreon\cite{A03}, Andreon et al.\cite{P4}). Therefore, 
the red sequence colour is a good redshift estimator, modulo
the poorly sampled region $0.6<z<0.9$. Recent spectroscopic observations
performed at VLT, as well optical photometry
at CTIO, by the XMM--LSS Survey have already filled this region.

\section*{Conclusions}

All clusters studied in Andreon et al.\cite{P4}, with one exception, are successfully colour--detected (Fig 1).
Most of the clusters are identified in X--rays, largely independent of the
optical luminosity and colour of the cluster galaxies.  Therefore, their
colour--detection is non--trivial. The majority of the clusters are
optically poor (Abell richness class 0 or lower) consistent with the low
computed X--ray luminosities and velocity dispersions. We have therefore
demonstrated, using real clusters, that a colour plus spatial
overdensity search technique can effectively identify optically poor
systems at intermediate to high redshifts (at least those previously
identified in X--rays). 

The colour
selection techniques provide a method to identify clusters possibly
sampling the cluster mass function deeper than (our) X--ray observations
(typically, 10 to 20 ks with XMM). 

The cluster redshift can be derived, with good accuracy,
from the colour of the red sequence (Fig 2). However, astrometry and
photometry do not constrain the extension in redshift of the identified
structure (a cluster--sized structure, or a much larger one?), and,
therefore, attention should be paied to systematics introduced by the
uncertain nature of such purely colour selected systems
in cosmological applications.
Ultimately, a spectroscopic survey of a random subsample of high redshift
colour--detected clusters is needed to ascertain the frequency of each
structure.

\end{document}